\documentclass[12pt,a4paper]{article}

\usepackage{amsmath}
\usepackage{amssymb}
\usepackage{tikz}
\usepackage{fixltx2e}
\usepackage{amsfonts}
\usepackage{graphicx}
\usepackage{subfig}
\usepackage{caption}
\usepackage{dcolumn}
\usepackage{bm}
\usepackage[T1]{fontenc}
\usepackage{textcomp}
\usepackage[
colorlinks,linkcolor=blue,citecolor=magenta]{hyperref}
\usepackage[top=3cm,right=2cm,bottom=2.5cm,left=2cm]{geometry}
\begin{document}
\title{Gauge generator for  bi-gravity and multi-gravity models}

\author{Ali Dokhani,$^{a,b}$\thanks{a.dokhani@ipm.ir}
Zahra Molaee,$^{b}$\thanks{zmolaee@ipm.ir} Ahmad Shirzad$^{a,b}$\thanks{shirzad@ipm.ir}\\
	{\it $^{a}$Department of Physics, Isfahan University of Technology, Isfahan, Iran}\\
	{\it$^{b}$School of Particles and Accelerators,}\\
	{\it Institute for Research in Fundamental Sciences (IPM),}\\
	{\it P.O.Box 19395-5531, Tehran, Iran}}
\maketitle

\begin{abstract}
	Following the Hamiltonian structure of bi-gravity and multi-gravity models in the full phase space, we have constructed the generating functional of diffeomorphism gauge symmetry. As is expected, this generator is constructed from the first class constraints of the system. We show that this gauge generator works well in giving the gauge transformations of the canonical variables.

\end{abstract}
\section{Introduction}
Modification of Einstein-Hilbert theory of general relativity is a great dream for almost one century. One direction in this regard is introducing a consistent covariant theory for massive gravity, beginning from the famous paper of Fierz and Pauli \cite{fp} and continuing with the important works of Van Dam, Veltman and Zakharov  \cite{wit2,tmg}, Vainshtein  \cite{ach} and Boulware-Deser  \cite{BD1}.

After almost 70 years, in 2010 the determinative paper of de Rham- Gabadadze and Tolley presented a special interaction term which leads to a massive gravity model \cite{dr} . Then Hassan and Rosen established that it is ghost free and lifted the model to one with arbitrary coordinates  where the flat metric is replaced by a background auxiliary tensor field $ f_{\mu\nu} $ \cite{HR1,mmg,mmg2}. Soon after, they added to the model a dynamical term for this tensor field. In this way, the massive gravity more or less was jointed to a theory with two tensor fields $ g_{\mu\nu} $ and $ f_{\mu\nu} $, i.e. the bi-gravity \cite{HR5}. 

The crucial point in all modified gravity models is absence of Boulware-Deser ghost. It is well-known that the best way to recognize the dynamical variables of a given theory in the non-linear level is the Hamiltonian analysis.  Several articles have been appeared on Hamiltonian analysis of massive gravity and bi-gravity\cite{HR2}-\cite{MS1}.  Despite  some challenges, it is finally established \cite{MS2,Hassan18} that HR bi-gravity possesses seven degrees of freedom corresponding to one massive and one massless graviton (and no ghost degree of freedom). 

 Extension of bi-gravity as multi-gravity models are also attractive theoretically \cite{sall}. It is interesting in description of some cosmological observations, as well \cite{s6}-\cite{s13}. However, there is no guarantee that the general form of the multi-gravity is ghost free except for sub classes of the type found in\cite{HS}.  Here we consider a special type of multi gravity which is a combination of bi-metric interactions \cite{HR5} for which the absence of ghosts is well-established. Explicit Hamiltonian analysis of this type of multi-gravity in the frame work of ADM variables  is  performed recently \cite{MS3}. 

Although, the main purpose in the literature for the Hamiltonian analysis of bi-gravity and multi-gravity theories has been counting the number of degrees of freedom, however, another reason for this investigation may be establishing the relationship between the constraint structure and the symmetries of the theories. As is well known, every gauge symmetry should be generated by the first class constraints of the system \cite{dirac}. In fact, the Hamiltonian analysis is not completed just by obtaining the expected number of degrees of freedom, i.e. a satisfactory analysis should also include the correct gauge transformations of the canonical variables via their Poisson brackets with a suitable generating functional \cite{gracia-pons,chai}. This generating functional of gauge transformations should be constructed from the first class constraints (of different levels of consistency process), as well as gauge parameters and their derivatives \cite{sh} and \cite{sh2}.

The problem of constructing the gauge generator is not an easy task, even for the simplest case of Einstein-Hilbert theory. This has been the subject of a sires of papers \cite{p1}-\cite{p3}. It turns out that the transformations due to diffeomorphism are not projectable, i.e.  the gauge generator can not be written directly in terms of the diffeomorphism parameters. However, it is illustrated \cite{Pons} that the arbitrary parameters in the gauge generator can be redefined in terms of the diffeomorphism parameters, where the relations depend on the dynamical variables. This algorithm can be followed for a special class of models where the canonical Hamiltonian is of the form $ H_{c}=\Sigma N_{\mu}H^{\mu} $. Fortunately, after imposing strongly the second class constraints, the canonical Hamiltonian of  bi-gravity and multi-gravity fall into this class.

In this paper, we use the above algorithm to construct the generating functional of gauge transformations for bi-gravity and multi-gravity. As we will see in the following sections, for both cases the variables $ N_{\mu} $ can be set as the lapse and shift functions of the reference metric $ f_{\mu \nu} $; so the above  procedure  goes on directly. The noticeable point is that both metrics should undergo similar diffeomorphism transformations with the same parameters. On the other hand, the first class constraints of the system contains mixtures of canonical variables of both metrics. Hence, it is instructive to have a single generating functional for gauge transformations of both metrics. We will show that our gauge generator gives the correct transformations for $ g_{ij} $, i.e. the spatial components of the second metric in bi-gravity. The same thing is true for the spacial part of component fields $g_{(k)ij}$ in multi-gravity.   However, the gauge transformations  of the dependent lapse and shift functions (components of $ g_{\mu\nu} $ or $g_{(k)\mu\nu}$'s) should be derived in some completed ways from the gauge transformations of canonical variables. 

In section 2 we will review the main features of the Hamiltonian structure of bi-gravity in ADM variables given in ref. \cite{MS2}. Then we use this structure in section 3 to establish the gauge generator of diffeomorphisms.  In section 4 we will consider a class of multi-gravity theories where a series of component metrics $g_{(k)\mu\nu}$ interact with a reference metric $f_{\mu \nu}$. After a brief review of the Hamiltonian structure, we will do the same thing for this model. Section 5 is devoted to our concluding remarks.

\section{ Hamiltonian structure of bi-gravity }
We start by introducing the HR bi-gravity model given by the following action\cite{mmg2},
\begin{equation}
S_{Bi-G}=M^{2}_{g} \int d^{4}x \sqrt{-g} 
\mathcal{R}(g)+M^{2}_{f}\int d^{4}x \sqrt{-f} 
\mathcal{R}(f)+2m^{4} \int d^{4}x \sqrt{-g}  \sum_{n=0}^{4} \beta_{n}e_{n}(\Bbbk). \label{a1}
\end{equation}
In Eq. (\ref{a1}) $\beta_{n}$ are free parameters, $m$ is a mass parameter and $M_{g}$ and $ M_{f}$ are two different Plank masses. The matrix  $ \Bbbk $ is  defined as $\Bbbk \equiv \sqrt{g^{-1}f}$ and $ e_{n}(\Bbbk) $ are elementary symmetric polynomials \cite{HR1}. 
 Here we impose for simplicity $ \beta_{0}=3, \beta_{1}=-1, \beta_2=\beta_3=0 $ and $ \beta_{4}=1 $ which come from a massive gravity with a flat reference metric. However, there are no such requirement in bi-gravity theory, and we can show that the results of this paper remains valid for every no zero value of $ \beta_{0}, \beta_{1}$ and $ \beta_{4} $.

  Decomposition of $ g_{\mu\nu} $ and $ f_{\mu\nu} $ in ADM approach are as follows
 \begin{equation}
 g_{\mu\nu}=
 \left( \begin{array}{cr}
 -N^{2}+N_{i}N^{i} & N_{i}  \\
 N_{i} &  g_{ij} \\
 \end{array}\right), \hspace{1cm}
 f_{\mu\nu}=
 \left(\begin{array}{cr}
 -M^{2}+M_{i}M^{i}       & M_{i}   \\
 M_{i}  & f_{ij}    \label{g1}  
 \end{array}\right).
 \end{equation}

By applying the following redefinition
\begin{equation}
N^{i}=M n^{i}+M^{i}+N D_{j}^{i}n^{j},\label{a15}
\end{equation}
for appropriate $3\times3$ matrix $D^{i}_{\ j}$ in which
 \begin{equation}
 D^{i}_{\ j}=\sqrt{g^{id}f_{dm}\mathcal{W}^{m}_{n}}(\mathcal{W}^{-1})^{n}_{j},\hspace{5mm}
\mathcal{W}^{l}_{\ j}=[1-n^{k}f_{km}n^{m}]\delta^{l}_{j}+n^{l}f_{mj}n^{m}.\label{g21}
 \end{equation}
the Lagrangian density would become linear in  lapses  $N$ and $ M$ and shifts  $M^{i}$ as follows \cite{mmg2}
\begin{eqnarray}&&
\mathcal{L}=M^{2}_{g}\pi^{ij}\partial_{t}g_{ij}+M^{2}_{f}p^{ij}\partial_{t}f_{ij}-M^{i} \mathcal{R}_{i} - M \mathcal{D} - N \mathcal{C}, \label{bi4}
\end{eqnarray}
where the momentum fields are defined as follow
\begin{eqnarray}&&
\pi^{ij}=-\sqrt{\gamma}(Z^{ij}-g^{ij}Z)\label{c1},\\&&
p^{ij}=-\sqrt{\gamma}(Y^{ij}-f^{ij}Y),\\&&
P_{M_{i}}\approx 0, P_{M}\approx 0,P_{N} \approx 0,P_{n^{i}}\approx 0,\label{bi9}
\end{eqnarray}
in which $Z^{ij}$ and $Y^{ij}$  are extrinsic curvatures on $g$ and $f$ metrics respectively. 
We also have
\begin{eqnarray}&&
\mathcal{C}=M^{2}_{g}\mathcal{R}_{0}^{g}+M^{2}_{g} D^{i}_{\ k} n^{k} \mathcal{R}_{i}^{g}-2m^{4}(\sqrt{g}\sqrt{x} D_{\ k}^{ k}-3\sqrt{g}),\label{bi57}\\ &&
\mathcal{D}=M^{2}_{f}\mathcal{R}_{0}^{f}+ M^{2}_{g} n^{i}\mathcal{R}_{i}^{g}-2m^{4}(\sqrt{g}\sqrt{x}-\sqrt{f}),\label{bi17}\\ && 
\mathcal{R}_{i}= M^{2}_{g} \mathcal{R}_{i}^{g}+M^{2}_{f}\mathcal{R}_{i}^{f},
\label{bi6}
\end{eqnarray}
where $ x=1-n^{i}f_{ij}n^{j}$.
The expressions $\mathcal{R}^{(g)}_{0}$ , $\mathcal{R}^{(g)}_{i}$   correspond  to the Hilbert-Einstein action of the metric $g_{\mu\nu}$, i.e. 
\begin{equation}
\mathcal{R}_{0}^{(g)}= M^{2}_{g}\sqrt{g}\mathcal{R}+\dfrac{1}{M^{2}_{g} \sqrt{g}}(\frac{1}{2}\pi^{2}-\pi^{ij}\pi_{ij}),\hspace{10mm} \mathcal{R}_{i}^{(g)}=2g_{ij} \triangledown _{k}(\pi^{jk}).\label{a14}
\end{equation}
We have similar expressions for  $\mathcal{R}^{(f)}_{0}$ and $\mathcal{R}^{(f)}_{i}$. 
The total Hamiltonian reads  
\begin{eqnarray} &&
\mathcal{H}_{T}=\mathcal{H}_{c}+uP_{N}+vP_{M}+u^{i}P_{M^{i}}+v^{i}P_{n^{i}}, \label{k111}
\end{eqnarray} 
in which  $ u,v,u_{i} $ and $v_{i}$ are 8 undetermined Lagrange multipliers. Consistency of the primary constraints $P_{M}$, $P_{N}$, $P_{M_{i}}$   leads to second level constraints $ \mathcal{C} $, $ \mathcal{D} $ and $ \mathcal{R}_{i} $ while consistency of $P_{n^{i}}$ gives 
\begin{eqnarray} &&
\lbrace P_{n^{i}},\mathcal{H}_{c}\rbrace \equiv -\mathcal{S}_{i} =
-\left(M \delta^{l}_{\ i}+N \frac{\partial(D^{l}_{\ j}n^{j})}{\partial n^{i}}\right)U_{l} \approx 0,\label{bi13}
\end{eqnarray}
where
\begin{equation}
U_{l}=M^{2}_{g}\mathcal{R}_{l}(g)-2m^{4}\sqrt{g}n^{k}f_{kj}\delta^{j}_{\ l}x^{-1/2} \approx 0.\label{bi15}
\end{equation}
 In this level, $ P_{n^{i}},U_{i}  $ are second class constraints and $ n^{k} $ would be determined by strongly imposing the constraint relation $ U_{l}=0 $. Consistency of  $\mathcal{R}_{i}  $ is satisfied identically. Hence,   $\mathcal{R}_{i}  $ and $ P_{M^{i}} $  are first class constraints.  Assuming $ \{\mathcal{C}, \mathcal{D}\}=\Gamma $, we see that the only physically acceptable result comes out \cite{MS2} on the sector $ \Gamma=0 $ of the phase space. So consistency of $ \Gamma $ gives 
\begin{eqnarray} &&
\Omega(x)\equiv \int d^{3}y \{\Gamma(x), \mathcal{H}_{c}(y)\}_{*}=E(x)M(x)+F(x)N(x),
\end{eqnarray}
where $\{\}_{*}$ means Dirac brackets and
\begin{eqnarray} &&
F(x)=\{\Gamma(x), \mathcal{C}(y)\}_{*}, \ 
E(x)=\{\Gamma(x), \mathcal{D}(y)\}_{*}
\end{eqnarray}
In the total Hamiltonian, one combination of the Lagrange multipliers $ u $ and $v$  would be obtained from 
consistency of $ \Omega $ and one other combination remain undetermined. Thus, we have four undetermined gauge parameters which should be related to diffeomorphism transformation. One may  changes the lapse variables to $ \bar{N}, M $ so that 
\begin{eqnarray} &&
\mathcal{H}_{c}=\bar{N}\mathcal{C}+M\mathcal{D}^{\prime}+M^{i}\mathcal{R}_{i}, \label{h1c}
\end{eqnarray}
where
\begin{eqnarray}&&
\bar{N}=N+\frac{E}{F} M, \label{nb}\\&&
 \mathcal{D}^{\prime}=\mathcal{D}-\frac{E}{F}\mathcal{C}.\label{nb1}
\end{eqnarray}
In this configuration we see that  $ \mathcal{D}^{\prime} $ and $ \mathcal{R}_{i} $ are first class constraints. On the other hand, consistency of $ \Gamma $ gives $ \Omega=\bar{N}F $ and  consistency of $ \Omega $ gives the Lagrange multiplier of   $ P_{\bar{N}} $.  

\section{Gauge generator for bi-gravity}
Now we are able to derive the generator of diffeomorphism for HR bi-gravity. To do this, we use the method given in ref. \cite{Pons} concerning a system with the canonical Hamiltonian
  $H=M^{\mu}H_{\mu}$ in which the momenta $ P_{\mu} $ conjugate to $ M^{\mu} $ are primary constraints. Assuming the secondary constraints $ H_{\mu} $ to be first class we may have
  \begin{equation}
  \left\lbrace H_{\mu},H_{\nu} \right\rbrace = C^{\sigma}_{\mu\nu}H_{\sigma},\label{kj}
  \end{equation}
 Then the generating functional of gauge transformations is proposed as 
  \begin{equation}
  G(t)= P_{\mu} \dot{\xi}^{\mu}+(H_{\mu}+M^{\rho}C^{\nu}_{\mu\rho}P_{\nu})\xi^{\mu},\label{Gen}
  \end{equation}
where  $ \xi^{\mu} $ are gauge parameters. The gauge variation of every physical variable $ \chi $ turns out to be $ \delta \chi=\{\chi, G(t)\} $.
Note that each pair of contracted indices in Eqs. (\ref{kj}) and (\ref{Gen}) and hereafter include a special integration as well.  For the special case of general relativity with the well-known Hilbert-Einstein action, i.e.
  \begin{equation}
  S_{HE}=\int d^{4}x [\dot{g}_{ij}\pi^{ij}-N\mathcal{R}^{0}(g)-N_{i}\mathcal{R}^{i}(g)],
  \end{equation}
  the generating functional (\ref{Gen}) leads to the standard form of general coordinate transformation 
  $x^{\mu}\longrightarrow x^{\mu}-\epsilon^{\mu}$, provided we assume
 \begin{equation}
 \epsilon^{0}=\frac{\xi^{0}}{M},\hspace{10mm} \epsilon^{i}=\xi^{i}-\xi^{0}\frac{M^{i}}{M}. \label{epsilon}
 \end{equation} 
  For the current case of bi-gravity, the final form of the action reads 
   \begin{equation}
S_{BG}=\int d^{4}x [\dot{g}_{ij}\pi^{ij}+\dot{f}_{ij}p^{ij}-\bar{N}\mathcal{C}-M\mathcal{D}^{\prime}-M^{i}\mathcal{R}_{i}].
  \end{equation}
Here we have
    \begin{equation}
 \mathcal{H}_{c}=M\mathcal{D}^{\prime}+M^{i}\mathcal{R}_{i}.
  \end{equation}
  where we have assumed all the second class constraints as strongly vanishing functions. In order to find the coefficients $ C^{\sigma}_{\mu\nu} $, let us consider the Poisson brackets among the first class constraints, i.e.
\begin{eqnarray} &&
\lbrace \mathcal{R}_{i}(x),\mathcal{R}_{j}(x)\rbrace=-\mathcal{R}_{j}(x)\partial_{x^{i}}\delta(x-y)+\mathcal{R}_{i}(y)\partial_{y^{j}}\delta(x-y),
\nonumber\\&&
\lbrace \mathcal{D}^{\prime}(x),\mathcal{R}_{i}(y)\rbrace=-\mathcal{D}^{\prime}(y)\partial_{x^{i}}\delta(x-y), \nonumber\\&&
\lbrace\mathcal{D}^{\prime}(x),\mathcal{D}^{\prime}(y)\rbrace=-f^{ij}(x)\mathcal{R}_{j}(x)\partial_{x^{i}}\delta(x-y)+f^{ij}(y)\mathcal{R}_{i}(y)\partial_{y^{j}}\delta(x-y).\label{rt}
\end{eqnarray}
 Comparing  Eqs. (\ref{kj}) and (\ref{rt}) shows that the coefficients $C^{\sigma}_{\mu\nu}$ are the  same as general relativity, i.e. 
\begin{eqnarray} &&
C^{i^{\prime \prime}}_{00^{\prime}}=f^{ij}(x^{\prime \prime})\left(\delta^{3}(x-x^{\prime\prime})+\delta^{3}(x^{\prime}-x^{\prime\prime})\right)\frac{\partial \delta^{3}(x-x^{\prime})}{\partial x^{j}},
\nonumber \\ &&
C^{0^{\prime \prime}}_{i0^{\prime}}=\delta^{3}(x-x^{\prime \prime})
\frac{\partial\delta^{3}(x-x^{\prime})}{\partial x^{i}}= -C^{0^{\prime \prime}}_{0^{\prime} i},
\nonumber \\ &&
C^{k^{\prime\prime}}_{i j^{\prime}}=\left(\delta^{k}_{i}\delta^{3}(x^{\prime\prime}-x^{\prime})\frac{\partial}{\partial x^{j}}+\delta^{k}_{j}\delta^{3}(x^{\prime\prime}-x)\frac{\partial}{\partial x^{i}}\right)\delta^{3}(x-x^{\prime}).
\end{eqnarray}
Inserting these  coefficients in Eq. (\ref{Gen}) we find  
\begin{eqnarray} &&
G(t) =\int d^{3}x P_{M}\dot{\xi}^{0}+P_{M^{i}}\dot{\xi}^{i}+\xi^{0}\mathcal{A}+\xi^{i}\mathcal{B}_{i}, \label{gen}
\end{eqnarray}
where
\begin{eqnarray} &&
\mathcal{A}=\mathcal{D}^{\prime}+P_{M^{i}}(x)f^{ij}(x) \partial_{j} M(x)+\partial_{j} \left( M(x)P_{M^{i}}(x)f^{ij}(x) \right)-
P_{M}(x)\partial_{i} M^{i}(x),\nonumber
\\&&
\mathcal{B}_{i}=\mathcal{R}_{i}+P_{M}(x) \partial_{i} M(x)+\partial_{j} \left(P_{M^{i}}(x)M^{j}(x)\right)+
P_{M^{j}}(x) \partial_{i} M^{j}(x).\label{genn}
\end{eqnarray}
Now the problem is  how we can relate the gauge parameters $ \xi^{\mu} $  to the diffeomorphism parameters $ \epsilon^{\mu} $. In this case, considering the reference metric $ f_{\mu\nu} $, we choose  relations (\ref{epsilon}) as introduced for General Relativity. 
As is well-known, the infinitesimal transformations due to diffeomorphism of the metric components are 
\begin{equation}
\delta f_{\mu\nu}=f_{\mu\nu,\rho}\epsilon^{\rho}+f_{\mu\rho}\epsilon_{,\nu}^{\rho}+f_{\rho\nu}\epsilon_{,\mu}^{\rho}, \label{gen4}
\end{equation}
In terms of the ADM variables, this corresponds to 
\begin{eqnarray}&&
\delta M= M_{,\mu}\epsilon^{\mu}+M\epsilon_{,0}^{0}-MM^{i}\epsilon^{0}_{,i}, \label{hg} \\ &&
\delta M^{i}= M^{i}_{,\mu}\epsilon^{\mu}+M^{i}\epsilon_{,0}^{0}-(M^{2}f^{ij}+M^{i}M^{j})\epsilon^{0}_{,j}+\epsilon^{i}_{,0}-M^{j}\epsilon^{i}_{,j},\\ &&
\delta f_{ij}= \epsilon^{0} \dot{f_{ij}}+2f_{ik} \partial_{j}\epsilon^{k}+2M_{i} \partial_{j}\epsilon^{0}+\epsilon^{e}\partial_{e}f_{ij}.\label{hg0}
\end{eqnarray}
Similar variations should be considered for $ N $, $ N_{i} $ and $ g_{ij} $ of $ g_{\mu\nu} $. 
The gauge variations $ M $, $ M_{i} $ and $ f_{ij} $ may be resulted directly under the Poisson brackets of the corresponding  variable with the generating functional $ G(t) $ as follows 
\begin{eqnarray}&&
\delta M=(\xi^{i}-\frac{M^i}{M}\xi^{0})\partial_{i}M+\partial_{0}\xi^{0}-M^{i}\partial_{i}\xi^{0}+\xi^{0}(M\partial_{0}(M^{-1})-MM^{i}\partial_{i}(M^{-1})+\frac{\partial_{0}M}{M}),\label{hj0}
\\&&
\delta M^{i}=-Mf^{ij}\partial_{j}\xi^{0}+\frac{\xi^{0}}{M}\partial_{0}M^{i}+\partial_{0}\xi^{i}-M^{j}\partial_{j}\xi^{i}+(\xi^{j}-\frac{M^{j}}{M}\xi^{0})\partial_{j}M^{i},
\\&&
\delta f_{ij}= \xi^{0}\lbrace f_{ij},\mathcal{D}^{\prime}\rbrace	+\xi^{i}\lbrace f_{ij},\mathcal{R}_{i}\rbrace. \label{hj1}
\end{eqnarray}
Note that in obtaining the Poisson brackets of $ M $, $ M_{i} $ and $ f_{ij} $ with $ G(t) $ we have considered terms containing $ P_{M} $, $ P_{M^{i}} $ and $ p_{ij} $ in the expressions (\ref{gen}) and (\ref{genn}), respectively. Then using Eqs. (\ref{nb1}) and (\ref{bi6}) for $ \mathcal{D}'$ and $ \mathcal{R}_{i} $ and then Eqs. (\ref{bi57}) and (\ref{bi17}) for $ \mathcal{D} $ and $ \mathcal{C} $ we have finally
\begin{equation}
\delta f_{ij}=\xi^{i}M^{2}_{f}\lbrace f_{ij},\mathcal{R}^{(f)}_{i}\rbrace+\xi^{0}M^{2}_{f}\lbrace f_{ij},\mathcal{R}^{(f)}_{0}\rbrace.
\end{equation}
This is the well-known  variations obtained for Einestain-Hilbert theory \cite{Pons}, as 
\begin{equation}
2M_{i}M^{2}_{f}\partial_{j}(\frac{\xi^{0}}{M})+M^{2}_{f}(\xi^{k}-\frac{M^{k}}{M}\xi^{0})\partial_{k}f_{ij}+\frac{\xi^{0}}{M}M^{2}_{f}\partial_{0}f_{ij}+2M^{2}_{f}f_{ik}\partial_{j}(\xi^{k}-\frac{M^{k}}{M}\xi^{0}).
\end{equation}
It is straightforward to see that under redefinitions (\ref{epsilon}) variations (\ref{hj0}-\ref{hj1}) reduce to standard variations (\ref{hg}-\ref{hg0}).
 We should also be sure about the gauge variations of the  variables $ N $, $ N_{i} $ and $ g_{ij} $. For the components $ g_{ij} $ we have
\begin{eqnarray}&&
\delta g_{ij}= \xi^{0}\lbrace g_{ij},\mathcal{D}^{\prime}\rbrace	+\xi^{i}\lbrace g_{ij},\mathcal{R}_{i}\rbrace.\label{hjj}
\end{eqnarray}
Again using Es. (\ref{nb1}), (\ref{bi57}) and (\ref{bi17}) for $\mathcal{D}^{\prime}  $ and Eq. (\ref{bi6}) for $\mathcal{R}_{i}  $ we have
\begin{equation}
\delta g_{ij}=\left(\xi^{0}M^{2}_{g}(n^{i}+\frac{N}{M} D_{j}^{i}n^{j})+\xi^{i}M^{2}_{g}\right)\lbrace g_{ij},\mathcal{R}^{(g)}_{i}\rbrace+\xi^{0}M^{2}_{g}\frac{N}{M}\lbrace g_{ij},\mathcal{R}^{(g)}_{0}\rbrace,
\end{equation}
where we have used the equality $ \frac{N}{M}=-\frac{E}{F} $ due to strongly vanishing of $ \bar{N} $ in Eq. (\ref{nb}). 
Using Eqs. (\ref{a15}),  (\ref{nb}) and (\ref{nb1}) we find the following result
\begin{eqnarray}&&
\delta g_{ij}=M^{2}_{g} \left(\xi^{0}\frac{N^{i}-M^{i}}{M}+\xi^{i}\right)\lbrace g_{ij},\mathcal{R}^{(g)}_{i}\rbrace+\xi^{0}M^{2}_{g}\frac{N}{M}\lbrace g_{ij},\mathcal{R}^{(g)}_{0}\rbrace.\label{hj}
\end{eqnarray}
Comparing Eq. (\ref{hj}) with Eq. (\ref{hj1}) for $ \delta f_{ij} $ we see that under similar combinations of the coefficients of the last two terms we have 
\begin{eqnarray}&&
\xi^{0}\frac{N^{i}-M^{i}}{M}+\xi^{i}-\frac{N^{i}}{N}(\frac{N }{M}\xi^{0})=\xi^{i}-\frac{M^{i}}{M}\xi^{0}
=\epsilon^{i}. \label{ui}
\end{eqnarray}
Hence, the same generating functional which gives $ \delta f_{ij} $ also results to correct variation for $ \delta g_{ij} $ with the same relationship between the gauge parameters $ \xi^{\mu} $ and the diffeomorphism parameter $ \epsilon^{\mu} $. So, for $ \delta g_{ij} $ we find the standard result 
\begin{eqnarray}&&
\delta g_{ij}= \epsilon^{0} \dot{g_{ij}}+2g_{ik} \partial_{j}\epsilon^{k}+2N_{i} \partial_{j}\epsilon^{0}+\epsilon^{e}\partial_{e}g_{ij}.
\end{eqnarray}
The generating functional $ G(t) $ should also give correct result for variations of $ N $ and $ N_{i} $ under diffeomorphism. However, $ \delta N $ and $ \delta N^{i} $ should be calculated indirectly in terms of the variations of other variables. Let us begin with the second class constraint $ \bar{N}=0 $ which implies 
\begin{eqnarray}&&
\delta  N=\delta (\frac{-E }{F}M).\label{ds}
\end{eqnarray}
As is seen we need to calculate $ \delta E $ and $ \delta F $ in terms of the variations of the canonical variables. 
In order to find variations 
 $ \delta N^{i} $ let us vary  Eq. (\ref{a15}) to find
\begin{eqnarray}&&
\delta N^{i}=\delta(M^{i}+M n^{i}+N D_{j}^{i}n^{j} ).\label{ej}
\end{eqnarray}
In this equation we should compute $ \delta n^{i} $ and $ \delta D^{i}_{j} $ which in turn depends on the canonical variables as well as $ n^{i} $. Remember that under imposing the constraints $ U^{i} $ in Eq. (\ref{bi15}) we can express $ n^{i} $ in terms of the canonical variables. Adding all these points together, we have a long way to calculate $ \delta n^{i} $ and $ \delta D^{i}_{j} $
in Eq. (\ref{ej}) as well as $ \delta E $ and $ \delta F $ in Eq. (\ref{ds}). 

These arguments show  that the variables $N$ and $N^i$ are not independent variables. So, their gauge transformations need not to be derived from particular expressions in the generating functional. In other words, there is no room to modify the gauge generator $ G(t) $ in order to get the gauge transformations of  $ \delta N $ and $ \delta N^{i} $. Hence, the only remaining task is to check that under the gauge variations of the canonical variables, the dependent expressions $ \delta N $ and $ \delta N^{i} $ comes out to have the correct form.  We have not done this explicitly, however, there is no reason that it may be violated. We will give more explanations about this point in the last section. 
\section{Gauge generator for multi-gravity}
Let us first review the Hamiltonian formalism of the multi-gravity model \cite{HR5} with $ \mathcal{N}-1 $ interacting component metrics $ g_{(k)\mu\nu} $ and one reference metric $  g_{(\mathcal{N})\mu\nu}\equiv f_{\mu\nu} $. The Lagrangian reads
\begin{equation}
S_{multi-G}=\int d^{4}x \left(\sum_{k=1}^{\mathcal{N}} M^{2}_{g_{(k)}}  \sqrt{-g_{(k)}} 
\mathcal{R}(g_{(k)})+2m^{4} \sqrt{-g_{(k)}}  \sum_{k=1}^{\mathcal{N}-1} \sum_{n=0}^{4} \beta^{(k)}_{n}e_{n}(\mathcal{K}^{(k)})\right), \label{bv1}
\end{equation}
where the matrix $\mathcal{K}^{(k)}$ is $ \sqrt{g_{(k)}^{-1}f}$,  $m$ is a mass parameter and $\beta^{(k)}_{n}$ are free parameters.
 We need not here to restrict ourselves to a special choice of parameters $\beta_n^{(k)}$. In fact, the details of expressions do depend on the values of $ \beta_{n}^{(k)} $. However, in order to indicate the general structure of gauge generating functional, it is not necessary to show the details of constraints, although they clearly depend on the parameters $\beta_{n}^{(k)}$. Hence, our analysis in this section is general, and is not limited to a special choice of the coefficient $ \beta_{n}^{(k)} $. 
As before, let us consider 
 the following  $ (\mathcal{N}-1)$ redefined shift variables 
\begin{equation}
N^{i}_{(k)}=M n^{i}_{(k)}+M^{i}+N_{(k)} D_{(k)j}^{i}n^{j}_{(k)}, \label{a25}
\end{equation}
where $ N^{i}_{(k)} $, $ N_{(k)} $, $ M $ and $ M^{i} $ are lapse and shift functions of $ g_{(k)\mu\nu} $ and $ f_{\mu\nu} $ respectively. Applying  relation (\ref{a25}), the action linearizes  versus all lapse variables as well as shift variables $M_i$. Hence, the canonical Hamiltonian takes the form 
\begin{equation}
\mathcal{H}_{c}=M \phi+N_{(k)} \phi_{(k)}+M^i \mathcal{R}_{i}. \label{hc}
\end{equation}
The momenta $P_{{(k)}}$, $P_{i(k)} $, $P$ and $P_{i}$   conjugate respectively to  $N_{(k)}$, $n^{i}_{(k)}$, $M$ and $ M^{i} $  are  primary constraints. 
Consistency of  primary constraints gives secondary constraints $\phi$, $ \phi^{(k)} $, $\mathcal{R}_{i}  $ and $ \mathcal{S}_{i(k)} $ where $\mathcal{R}_{i}= \sum \mathcal{R}_{i}^{(\mathcal{N})} $ and  $ \mathcal{S}_{i(k)} $ are analogous to Eq. (\ref{bi13}). 
Direct calculation shows that $\{\phi_{(k)},\phi_{(k')}\}\approx 0$ for all $k$ and $k'$, and the only non-vanishing Poisson brackets among second level constraints are $\psi_k=\{\phi,\phi_{(k)}\}$ (see ref. \cite{MS3}). The physics of the system proceeds correctly if we restrict the dynamics to the subspace defined by the new constraint $ \psi_k $. This implies  consistency of $ \phi_{(k)} $'s are satisfied identically.

To proceed  we should  consider  consistency of  $ \psi_{k} $'s. Assuming $\mathcal{G}_{k'k} \equiv \{\psi_{(k')},\phi_{(k)}\}$  and $\mathcal{G}_{k^{\prime}} \equiv \{\psi_{(k')},\phi\}$, this implies
\begin{eqnarray} &&
N_{(k)}=-\mathcal{G}_{k'k}^{-1}\ \mathcal{G}_{k^{\prime}}
M.
\end{eqnarray}
Defining the modified lapse functions 
\begin{eqnarray} &&
\bar{N}_{(k)} =N_{(k)}+\mathcal{G}_{k'k}^{-1}\ \mathcal{G}_{k^{\prime}} M, \label{zg0}
\end{eqnarray}
and assuming 
\begin{eqnarray} &&
\phi^{\prime}=\phi-\mathcal{G}_{k^{\prime}}(\mathcal{G}^{-1^{T}})_{k'k}\ \  \phi_{(k)}M,\label{zf1}
\end{eqnarray}
the canonical Hamiltonian reads 
\begin{eqnarray} &&
\mathcal{H}_{c}=\bar{N}_{(k)}\phi_{(k)} +M\phi^{\prime}+M^{i} \mathcal{R}_{i}.\label{zu3}
\end{eqnarray}
Hence, consistency of $ \psi_{k} $'s gives the last level constraints $ \bar{N}_{(k)}\approx 0 $  which are second class with their corresponding momenta $ \bar{P}_{(k)}$. 
 
In this way, we have  $ 8 $  first class constraints ($ P,P_{i},\mathcal{R}_{i}, \phi^{\prime} $) for generating the space-time diffeomorphism. It
 turns out that there are $ (\mathcal{N}-1)\times 6  $ second class constraints $ (\mathcal{S}^{i}_{(k)},P_{i(k)}) $ 
and   $ (\mathcal{N}-1) \times 4$ second class constraints $\bar{N}_{(k)},\psi_{(k)}, \phi_{(k)}$ and $\bar{P}_{(k)}   $. This corresponds to $ 2 \times (5 \mathcal{N}-3) $ dynamical degrees of freedom which describes a system with $\mathcal{N}-1  $ massive gravitons and one massless graviton.

Now let us go through constructing of the gauge generator for multi-gravity system. Imposing strong equalities for vanishing the second class constraints
\begin{eqnarray} &&
\bar{N}_{(k)} =N_{(k)}+\mathcal{G}_{k'k}^{-1}\ \mathcal{G}_{k^{\prime}} M=0, \label{po}
\end{eqnarray}
the canonical Hamiltonian reads 
\begin{eqnarray} &&
\mathcal{H}_{c}=M\phi^{\prime}+M^{i} \mathcal{R}_{i}.\label{z3}
\end{eqnarray}
 The gauge generator is the same as Eq. (\ref{Gen}),
where the coefficient $C^{\sigma}_{\mu\nu}$  should be derived from the Poisson brackets of constraints $ \phi' $ and $ \mathcal{R}^{i} $ as follows
\begin{eqnarray} &&
\lbrace \mathcal{R}_{i}(x),\mathcal{R}_{j}(x)\rbrace=-\mathcal{R}_{j}(x)\partial_{x^{i}}\delta(x-y)+\mathcal{R}_{i}(y)\partial_{y^{j}}\delta(x-y),
\nonumber\\&&
\lbrace \phi^{\prime}(x),\mathcal{R}_{i}(y)\rbrace=-\phi^{\prime}(y)\partial_{x^{i}}\delta(x-y), \nonumber\\&&
\lbrace\phi^{\prime}(x),\phi^{\prime}(y)\rbrace=-f^{ij}(x)\mathcal{R}_{j}(x)\partial_{x^{i}}\delta(x-y)+f^{ij}(y)\mathcal{R}_{i}(y)\partial_{y^{j}}\delta(x-y).\label{tr}
\end{eqnarray}
One can read from Eq. (\ref{tr}) 
\begin{eqnarray} &&
C^{i^{\prime \prime}}_{00^{\prime}}=h^{ij}(x^{\prime \prime})\left(\delta^{3}(x-x^{\prime\prime})+\delta^{3}(x^{\prime}-x^{\prime\prime})\right)\frac{\partial \delta^{3}(x-x^{\prime})}{\partial x^{j}},
\nonumber \\ &&
C^{0^{\prime \prime}}_{i0^{\prime}}=\delta^{3}(x-x^{\prime \prime})
\frac{\partial\delta^{3}(x-x^{\prime})}{\partial x^{i}}= -C^{0^{\prime \prime}}_{0^{\prime} i},
\nonumber \\ &&
C^{k^{\prime\prime}}_{i j^{\prime}}=\left(\delta^{k}_{i}\delta^{3}(x^{\prime\prime}-x^{\prime})\frac{\partial}{\partial x^{j}}+\delta^{k}_{j}\delta^{3}(x^{\prime\prime}-x)\frac{\partial}{\partial x^{i}}\right)\delta^{3}(x-x^{\prime}).
\end{eqnarray}
Inserting these coefficients in Eq. (\ref{Gen}) we find 
\begin{eqnarray} &&
G(t) =\int d^{3}x P\dot{\xi}^{0}+P_{i}\dot{\xi}^{i}+\xi^{0} \mathcal{A}+\xi^{i} \mathcal{B}_{i},\label{gen1}
\end{eqnarray}
where
\begin{eqnarray} &&
\mathcal{A}=\phi'+
P_{i}(x) f^{ij}(x) \partial_{j} M(x)+ \partial_{j} \left( M(x)P_{i}(x)f^{ij}(x) \right)
- P(x) \partial_{i} M^{i}(x) 
\nonumber \\ &&
\mathcal{B}_{i}=\mathcal{R}_{i}+ P(x) \partial_{i} M(x)
+  \partial_{j} \left(P_{i}(x)M^{j}(x)\right)+ P_{j}(x)  \partial_{i} M^{j}(x). \label{gen11}
\end{eqnarray}
In order to see whether the gauge generator (\ref{gen1}) works well, we first calculate the gauge variations of the components of the reference metric $ f_{\mu\nu} $.  Using the  relations (\ref{epsilon}) between the gauge parameters $ \xi^{\mu} $ and diffeomorphism parameters $ \epsilon^{\mu} $ the result is the same as given in Eqs. (\ref{hj1}), (\ref{hjj}) and (\ref{hj}). Then let us calculate the gauge transformations of the spatial part of the component metrics $  g_{(k)ij}  $ as follows
\begin{eqnarray}&&
\delta g_{(k)ij}=\xi^{0}\lbrace g_{(k)ij},\phi^{\prime}\rbrace	+\xi^{i}\lbrace g_{(k)ij},\mathcal{R}_{i}\rbrace=\nonumber\\&&(\xi^{0}M^{2}_{g}(n_{(k)}^{i}+\frac{N_{(k)}}{M} D_{(k)j}^{i}n_{(k)}^{j})+\xi^{i}M^{2}_{g})\lbrace g_{(k)ij},\mathcal{R}^{g_{(k)}}_{i}\rbrace+\xi^{0}M^{2}_{g}\frac{N_{(k)}}{M}\lbrace g_{(k)ij},\mathcal{R}^{g_{(k)}}_{0}\rbrace\nonumber\\&&
=M^{2}_{g} (\xi^{0}\frac{N_{(k)}^{i}-M^{i}}{M}+\xi^{i})\lbrace g_{(k)ij},\mathcal{R}^{g_{(k)}}_{i}\rbrace+\xi^{0}M^{2}_{g}\frac{N_{(k)}}{M}\lbrace g_{(k)ij},\mathcal{R}^{g_{(k)}}_{0}\rbrace. 
\end{eqnarray}
It is not difficult to check that under redefinition (\ref{ui}) the variations $ \delta g_{(k)ij}  $ take the standard form (similar to Eq. (\ref{hg0})). Hence, the same generating functional gives simultaneously the diffeomorphism transformation of all $ g_{(k)ij} $'s, as well as the reference metric $ f_{ij} $. However, we may wish to derive the gauge variation  
$ \delta N_{(k)}$ and $\delta N_{(k)}^{i}  $. As mentioned in the last paragraph of the previous section, the variables $ N_{(k)} $ and $ N_{(k)}^{i} $ are dependent to other variables through Eqs. (\ref{a25}) and (\ref{po}), so we have 
\begin{equation}
\delta N^{i}_{(k)}=\delta (M n^{i}_{(k)}+M^{i}+M_{(k)} D_{(k)j}^{i}n^{j}_{(k)}), \label{a115}
\end{equation}

\begin{eqnarray} &&
\delta N_{(k)}=\delta (\mathcal{G}_{k'k}^{-1}\ \mathcal{G}_{k^{\prime}} M). \label{z0}
\end{eqnarray}
Then one needs to take into account the equations defining $\mathcal{G}_{k^{\prime}} $, $ \mathcal{G}_{k'k}^{-1} $ and $D_{(k)j}^{i}$ in order to find their  variations in terms of the canonical variables. Therefore, similar to the case of bi-gravity these calculations are too lengthy (see our discussions in the last section).

\section{Conclusions }
The main goal of this paper is completing the Hamiltonian analysis of a series of modified gravities which contain one or more massive gravitons  together with one single massless graviton. Despite some exceptions (see i.g. ref \cite{Hassan18}), the main focus in the literature is mostly on investigating  the existence or absence of the Boulware-Deser ghost through counting the dynamical variables in the Hamiltonian framework.

However, the Hamiltonian analysis has more capacities than this simple task. Especially, the Hamiltonian structure of a given model may help us to investigate the gauge symmetries of the system. This goal is achieved through constructing the generating functional of gauge transformations by using the first class constraints of the system. 

For general relativity and all its covariant extensions, the main gauge symmetry is the diffeomorphism which contains four infinitesimal arbitrary  fields. Hence, the constraint structure should necessarily contain a multiple of four first class constraints. Of curse, this should be the case if we take into account all the components of the metric (or metrics), i.e. we should consider the full phase space of the theory which include the lapse and shift functions and their corresponding momenta. Hence, those analysis which omit lapses and shifts in advance, or consider them from the very beginning as Lagrange multipliers, are not capable to recognize precisely the needed first class constraint which generate the gauge symmetry. In general, construction of the gauge generating functional is not an easy task. In ref. \cite{zms2} we can find instructions for doing this. The problem is even more complicated for general covariant theories with diffeomorphism as the gauge 
symmetry of the system. 

Fortunately,  the algorithm given by \cite{Pons} was capable to solve our problem here. We found suitable forms for gauge generators  of bi-gravity and multi-gravity which give correct gauge transformations for all spatial parts of the component metrics as well as the reference metric. However, as  a consequence of the Hamiltonian analysis, lapses and shifts of the component metrics turn out to be dependent variables. Therefore, there exist clear and straightforward instructions to obtain gauge variations of these lapses and shifts, though there should be done a lot of cumbersome calculations. However, since the gauge symmetry is clearly known from a covariant Lagrangian observation, there is no reason to be  in doubt about the result. We think it is just satisfactory to have a gauge generating functional which gives the corrects gauge transformations for spatial components of the metrics which constitute the canonical variables of the system.

 \end{document}